\documentclass[twocolumn,showpacs,preprintnumbers,amsmath,amssymb]{revtex4}

\usepackage{graphicx}
\usepackage{dcolumn}
\usepackage{bm}

\usepackage{epsfig}

\newcommand{\be}{\begin{equation}}\newcommand{\ee}{\end{equation}}
\newcommand{\bea}{\begin{eqnarray}}\newcommand{\eea}{\end{eqnarray}}
\newcommand{\brr}{\begin{array}}\newcommand{\err}{\end{array}}
\newcommand{\bit}{\begin{itemize}}\newcommand{\eit}{\end{itemize}}
\newcommand{\ben}{\begin{enumerate}}\newcommand{\een}{\end{enumerate}}

\newcommand{\ba}{\begin{array}}
\newcommand{\ea}{\end{array}}

\def\lab{\label}\def\lan{\langle}
\def\lf{\left}
\def\Lrar{\Leftrightarrow}
\def\non{\nonumber}\def\pa{\partial}\def\ran{\rangle}

\def\ri{\right}\def\wti{\widetilde}
\def\al{\alpha}\def\bt{\beta}
\def\de{\delta}\def\ep{\epsilon}
\def\te{\theta}
\def\si{\sigma}
\def\om{\omega}
\newcommand{\mlab}[1]{\label{#1}}

\def\1{{_{1}}}\def\2{{_{2}}}
\def\bp{{\bf {p}}}\def\bk{{\bf {k}}}
\def\bx{{\bf {x}}}
\def\bq{{\bf {q}}}

\newcommand{\ide}{1\hspace{-1mm}{\rm I}}
\newcommand{\noH}{:\;\!\!\;\!\!:H:\;\!\!\;\!\!:}
\def\noHe0{:\;\!\!\;\!\!:H_e(0):\;\!\!\;\!\!:}
\def\noHm0{:\;\!\!\;\!\!:H_\mu(0):\;\!\!\;\!\!:}
\begin{document}

\title{Lorentz invariance for mixed neutrinos}

\author{Massimo Blasone}
\email{blasone@sa.infn.it}
\affiliation{Dipartimento di
Fisica and INFN, Universit\`a di Salerno, 84081 Baronissi (SA), Italy}

\author{Jo\~{a}o Magueijo}
\email{j.magueijo@ic.ac.uk}
\affiliation{The Blackett Laboratory, Imperial College London, London
SW7 2AZ, U.K.}

\author{Paulo Pires Pacheco}
\email{p.pires-pacheco@ic.ac.uk}
\affiliation{The Blackett Laboratory, Imperial College London, London
SW7 2AZ, U.K.}


\begin{abstract}
We show that a proper field theoretical treatment of mixed (Dirac) neutrinos
leads to  non-trivial dispersion relations for the flavor states. We analyze such a
situation in the framework of the non-linear relativity schemes  recently proposed
by Magueijo and Smolin. We finally examine the  experimental implications of our theoretical
proposals by considering the spectrum and the end-point of beta decay in tritium.
\end{abstract}

\pacs{14.60.Pq,11.30.Cp,03.70.+k}

\maketitle

\section{Introduction}

The subject of neutrino oscillations has now matured from an
insightful prediction by Bruno Pontecorvo \cite{Pont} and the
early results of Homestake \cite{homestake} to a structured
framework backed by a wealth of new quantitative  data
\cite{Kam,SNO,kamland,K2K}. This advances have been paralleled by
much progress on the theoretical front with the efforts divided
between phenomenological pursuits
 of more refined oscillation formulas and attempts
to give the theory a sound formal structure within Quantum Field Theory (QFT).

A major outstanding  question was that of the
existence of a  Hilbert space for the flavour states \cite{Fujiiold}. The
Pontecorvo treatment of the latter in Quantum Mechanics (QM)  actually
turns out to be forbidden by the Bargmann super-selection rules
\cite{Barg}. This naturally pointed to QFT where the problem found
its resolution \cite{BV95,BHV99,hannabuss,fujii}. Subsequently an
even more consistent picture emerged with the discovery of an
associated geometric phase \cite{Berry}, the extension to the case
of three-flavours \cite{3flav} and bosons \cite{bosonmix,Ji} and
to the case of neutral fields \cite{BP03}. The study of
relativistic flavor currents \cite{currents,Blasone:2002wp} was recently used to
solve the phenomenologically very relevant problem of finding a
space-oscillation formula \cite{Beuthe}.

Another important outcome of these studies is the understanding that
the flavour eigenstates constitute the real
physical entities, in contrast with the common view where the mass eigenstates
are taken to be the fundamental objects \cite{giunti}.

The present paper proceeds in that direction by finding dispersion
relations for the mixed neutrinos taking into account their nature
as fundamental particles. We find that these dispersion relations
no longer have the standard form thus exhibiting some form of
breakdown of Lorentz invariance. This development is rather timely
given the strong interest generated by various schemes involving
such modifications
\cite{Moffat:2002zy,Albrecht:1998ir,Alexander:2001ck,lorentz,Kimberly:2003hp,Amelino-Camelia:2000mn}

We further study the experimental implications of our analysis and compare it
with the standard treatment,
by considering the various possibilities which can arise in the
 end-point of
the beta decay of tritium depending on which scenario turns out to be true.\\

The paper is organized as follows: In section II we show how flavor
states can be properly defined in QFT.
In section III we then consider the dispersion relations
associated to such states. In section IV
we study the covariance of these forms and  the
description of the non-linear representation of the Poincar\'e
algebra necessary to support them.  Finally in section VI
we propose experimental tests, with special emphasis on the
end-point of beta decay of tritium. Section VII is devoted to
conclusions.

\section{Flavor neutrino states in Quantum Field Theory}\label{relcur}

Let us begin our discussion by considering the following
Lagrangian density describing two free Dirac fields  with a mixed mass
term (see Appendix for conventions and further details):
\bea\label{lagemu} {\cal L}(x)\,=\,  {\bar \Psi_f}(x) \lf( i
\not\!\partial -
  M \ri) \Psi_f(x)\, ,
\eea
where $\Psi_f^T=(\nu_e,\nu_\mu)$ and $M = \lf(\brr{cc} m_e &
m_{e\mu} \\ m_{e\mu} & m_\mu \err \ri)$. The mixing
transformations
\bea\non \nu_e(x) & = &\cos\theta \,\,\nu_1(x) + \sin\theta\,\,
\nu_2(x)
\\[2mm] \label{fermix}
\nu_{\mu}(x) & =& - \sin\theta\,\, \nu_1(x) +
\cos\theta\,\,\nu_2(x)   \eea
with $\te$ being the mixing angle, diagonalize the quadratic form
of Eq.(\ref{lagemu}) to the Lagrangian for two free Dirac fields,
with masses $m_1$ and $m_2$:
\bea\label{lag12} {\cal L}(x)\,=\,  {\bar \Psi_m}(x) \lf( i
\not\!\partial -
  M_d\ri) \Psi_m(x)  \, ,
\eea where $\Psi_m^T=(\nu_1,\nu_2)$ and $M_d = diag(m_1,m_2)$. One
also has $ m_{e} = m_{1}\cos^{2}\te + m_{2} \sin^{2}\te~$,
$m_{\mu} = m_{1}\sin^{2}\te + m_{2} \cos^{2}\te~$, $m_{e\mu}
=(m_{2}-m_{1})\sin\te \cos\te\,$. Without loss of generality we
take $\te$ ranging from $0$ to $\frac{\pi}{4}$ (maximal mixing)
 and $m_2> m_1$.

The generator for the mixing relations (\ref{fermix}) can be
introduced as \cite{BV95}:
\bea \nu_\si(x) & \equiv & G^{-1}_\te(t) \, \nu_j(x) \, G_\te(t)
\, ,
\\ [2mm]
G_{\te}(t)& =& exp\lf[\te \int d^{3}{\bf x}  \lf(\nu_{1}^{\dag}(x)
\nu_{2}(x) - \nu_{2}^{\dag}(x) \nu_{1}(x) \ri)\ri] \eea
with $(\si,j)=(e,1) , (\mu,2)$ and $t\equiv x_0$.
Note that  $G_\te(t)$ does not
leave invariant the vacuum $|0 \ran_{1,2}$:
\bea \lab{flavac} |0 (t)\ran_{e,\mu} = G^{-1}_\te(t)\; |0
\ran_{1,2}\;. \eea
We will refer to $|0(t) \ran_{e,\mu}$ as to the {\em flavor vacuum}: it is
orthogonal to $|0 \ran_{1,2}$ in the infinite volume
limit \cite{BV95}. We define the flavor annihilators, relative to
the fields $\nu_{e}(x)$ and $\nu_{\mu}(x)$ as
\bea \label{ladder}\al^{r}_{{\bf k},\si}(t) &\equiv&
G^{-1}_{\te}(t)\,\al^{r}_{{\bf k},j} \,G_{\te}(t)\\ [2mm]
\bt^{r\dag}_{{-\bf k},\si}(t)&\equiv&
 G^{-1}_{\te}(t) \, \bt^{r\dag}_{{-\bf k},j}\,
G_{\te}(t) \eea with $(\si,j)=(e,1) , (\mu,2)$. The flavor fields
can then be expanded in analogy to the free field case:
\bea  \non  \nu_\si(x)&=& \sum_{r=1,2} \int \frac{d^3
\bk}{(2\pi)^\frac{3}{2}}\, \lf[ u^{r}_{{\bf k},j}(t) \,\al^{r}_{{\bf
k},\si}(t)\ri.
\\ \label{flavfields} && \lf.
+   \, v^{r}_{-{\bf k},j}(t)\, \bt^{r\dag}_{-{\bf
k},\si}(t) \ri] \,e^{i {\bf k}\cdot{\bf x}}\,, \eea
with $(\si,j)=(e,1) , (\mu,2)$.

The symmetry properties of the Lagrangian (\ref{lagemu}) have been
studied in Ref.\cite{currents}: one has a total conserved charge $Q$
associated
with the global $U(1)$ symmetry and time-dependent charges associated to the
(broken) $SU(2)$ symmetry. Such charges are the relevant physical quantities
for the study of flavor oscillations \cite{BHV99,3flav}. They are also
essential in the definition of (physical) flavor neutrino states, as the
one produced in a beta decay, for example.

In the present case of two flavors, we obtain for the flavor charges \cite{currents}:
\bea
& & Q_\si(t)  = \int d^3 \bx  \;\nu^\dag_\si(x) \,\nu_\si(x)\,
\\ \non&&= \, \sum_{r} \int d^3 \bk \, \lf( \al^{r\dag}_{{\bf k},\si}(t)
\al^{r}_{{\bf k},\si}(t)\, -\, \bt^{r\dag}_{-{\bf
k},\si}(t)\bt^{r}_{-{\bf k},\si}(t)\ri)\,,  \eea
with $\sigma = e,\mu$.
By indicating with $Q_j$ ($j=1,2$), the (conserved) charge  operators for the
free fields, we obtain the following relations:
\bea\lab{Qeq} && \hspace {-.3cm} Q_\sigma(t)= G_\theta^{-1}(t)  Q_j
G_\theta(t)\; ,\;  \,(\sigma,j) = (e,1),(\mu,2),
\\ [3mm]\lab{HQtot}
&&\hspace {-.3cm} \sum_\si  Q_\si(t) =
\sum_j  Q_j =  Q\quad;\quad [Q,G_\theta(t)]
\neq 0 \eea
Thus the single neutrino and antineutrino states of definite
flavor are defined in the following way:
\bea \non &&Q_{\sigma}(t)\,|\nu_{\sigma}^{\bf k}(t)\ran \,=\,
|\nu_{\sigma}^{\bf k}(t)\ran ,
\\ [2mm]\label{charge}
&& Q_{\sigma}(t)\,|{\bar \nu}_{\sigma}^{\bf k}(t)\ran \,=\,
- |{\bar \nu}_{\sigma}^{\bf k}(t)\ran
\eea
and  they naturally turn out to be vectors
of the flavor Hilbert space ${\cal H}_{e,\mu}$:
\bea \label{flasta1}
&&|\nu_{\sigma}^{\bf k}(t)\ran = \al^{r\dag}_{{\bf
k},\si}(t)\;|0(t)\ran_{e,\mu}\;
\\ [2mm] \label{flasta2}
&& |{\bar \nu}_{\sigma}^{\bf k}(t)\ran = \bt^{r\dag}_{{\bf
k},\si}(t)\;|0(t)\ran_{e,\mu} \eea

One can also define the momentum operator for mixed fields \cite{BP03}:
\bea \label{momop}&&{\bf P}_\sigma(t) =
\int d^3{\bf x}\, \nu^\dagger_\sigma(x)
(-i\nabla)\nu_\sigma(x)
\\ \non&&= \sum_r\int d^3 {\bf k}\,\,  {\bf k}
\left( \alpha^{r\dagger}_{{\bf k},\sigma}(t) \alpha^{r}_{{\bf
k},\sigma}(t) + \beta^{r\dagger}_{{\bf
k},\sigma}(t)\beta^{r}_{{\bf k},\sigma}(t) \right) \eea
with
\bea\lab{Peq} && \hspace {-.3cm} {\bf P}_\sigma(t)= G_\theta^{-1}(t)  {\bf P}_j
G_\theta(t)\; ,\;  \,(\sigma,j) = (e,1),(\mu,2),
\\ [3mm]\lab{HQtot}
&&\hspace {-.3cm} \sum_\si  {\bf P}_\si(t) =
\sum_j  {\bf P}_j =  {\bf P}\quad;\quad [{\bf P},G_\theta(t)]
\neq 0 \eea
where ${\bf P}_j$ ($j=1,2$) are the (conserved) momentum  operators for the
free fields and ${\bf P}$ is
the total  momentum operator for the system (\ref{lagemu}), (\ref{lag12}).
It is immediate to verify that the flavor states
Eq.(\ref{flasta1}),(\ref{flasta2}) have definite momentum (and helicity):
\bea
 &&{\bf P}_{\sigma}(t)\,|\nu_{\sigma}^{\bf k}(t)\ran \, =\, \bk
\,|\nu_{\sigma}^{\bf k}(t)\ran \\ [2mm] \label{momentum}
 && {\bf P}_{\sigma}(t)\,|{\bar \nu}_{\sigma}^{\bf k}(t)\ran \,=\,
 \bk \,|{\bar \nu}_{\sigma}^{\bf k}(t)\ran.
 \eea

\vspace{.2cm}

Note that the above defined flavor states differ from the ones
commonly used which are defined by (erroneously) assuming that the
Hilbert spaces for the flavor and the mass fields are the same.
For further convenience we denote with a index ``P'' the
Pontecorvo flavor states:
\bea  \non |\nu_e\rangle_P & = &\cos\theta \,\,|\nu_1\rangle +
\sin\theta\,\, |\nu_2\rangle
\\ [2mm] \label{Pont}
|\nu_{\mu}\rangle_P  & =& - \sin\theta\,\, |\nu_1\rangle  +
\cos\theta\,\,|\nu_2\rangle \eea
for which we do not specify the momentum index: as it is well
known \cite{Giunti:2001kj}, the flavor states so defined cannot
have the same momentum or energy in all inertial frames. Note also
that the  states (\ref{Pont}) are not eigenstates of the momentum
and charge operators (defined in Eqs.(\ref{charge}) and
(\ref{momop})) as they are not the vectors of the flavor
Hilbert space.

In the following, we will work in the Heisenberg picture, so the
Hilbert space is chosen at the reference time $t=0$. We thus
define our flavor states like
\bea \non &&|\nu^{\bf k}_{\sigma}\ran \equiv\; \al^{r\dag}_{{\bf
k},\si}(0)\;|0(0)\ran_{e,\mu}\,,
\\[2mm] && |{\bar \nu}^{\bf
k}_{\sigma}\ran \equiv\;\bt^{r\dag}_{{\bf
k},\si}(0)\;|0(0)\ran_{e,\mu}\,, \quad \sigma = e,\mu\;. \eea

\section{Dispersion relations for mixed neutrinos}

Let us now consider the explicit expression for the one
electron-neutrino state with definite helicity and momentum (at $t
=0)$:
\bea  \non|\nu^\bk_e \rangle  & =&  \prod_{\bk}
G^{-1}_{\bk}(\te)\, \alpha_{{\bk},1}^{r\dag}\, |0\rangle_{1,2}
= \left[\cos\theta\,\alpha_{{\bk},1}^{r\dag} +\ri.
\\ \non
& + & \lf. |U_{\bk}|\;
\sin\theta\;\alpha_{{\bk},2}^{r\dag}\ri.
\lf.-\epsilon \; |V_{\bk}|
\,\sin\theta \;
\alpha_{{\bk},1}^{r\dag}\alpha_{{\bk},2}^{r\dag}\beta_{{\bk},1}^{r\dag}
 \right]\times
\\  \label{one}
&\times& G^{-1}_{\bk,s\neq r}(\te)\prod_{\bp \neq \bk}
 \,G^{-1}_{\bp}(\te)\,|0\rangle_{1,2} \,.
\eea
where we used $G_\te(t)=\prod_{\bk} G_{\bk}(\te)$. Eq.(\ref{one})
shows clearly the non-trivial condensate structure of the flavor neutrino
states in terms of the mass eigenstates.

Next we  consider the energy-momentum tensor.  For the
massive fields $\nu_j$ we have:
\bea
{\cal J}^{\mu\nu}_j(x) \equiv i\,
\overline{\nu}_j(x) \,\gamma^\nu \partial_\mu\,  \nu_j(x) \quad ,\quad j = 1,2
\eea
from which the Hamiltonians for the free fields  $\nu_1$, $\nu_2$ trivially
follow:
\bea \non
&&H_j= i \int d^3{\bf x} \,\nu_j^\dag(x)\,\pa_0 \nu_j(x)
\\ \non
&&= \sum_r\int d^3{\bf k}\,   \left( \alpha^{r\dagger}_{{\bf
k},j}(t) \,\pa_0\, \alpha^{r}_{{\bf k},j}(t) + \beta^{r}_{{\bf
{k}},j}(t) \,\pa_0\, \beta^{r\dagger}_{{\bf k},j}(t) \right)
\\ \label{H12}
&&= \sum_r\int d^3{\bf k}\,  \om_{k,j} \left(
\alpha^{r\dagger}_{{\bf k},j} \, \alpha^{r}_{{\bf k},j} -
\beta^{r}_{{\bf {k}},j} \, \beta^{r\dagger}_{{\bf k},j} \right)
\; ,\;  \eea
with $j = 1,2.$ In a similar way we define the energy-momentum tensor for the
flavor fields:
\bea {\cal J}^{\mu\nu}_\si(x) \equiv i\,
\overline{\nu}_\si(x)\,\gamma^\nu \partial_\mu \, \nu_\si(x)
\quad ,\quad\si = e,\mu. \eea
The energy operators are now:
\bea \lab{Hflav} && H_\si(t)= i \int d^3{\bf x} \,\nu_\si^\dag(x)\,\pa_0
\nu_\si(x)
\\ \non
&&= \sum_r\int d^3{\bf k}\,   \left( \alpha^{r\dagger}_{{\bf
k},\si}(t) \,\pa_0\, \alpha^{r}_{{\bf k},\si}(t) + \beta^{r}_{{\bf
{k}},\si}(t) \,\pa_0\, \beta^{r\dagger}_{{\bf k},\si}(t) \right)
\eea
with $\si = e,\mu$. We also easily recover the momentum operators
(\ref{momop}). Notice that Eq.(\ref{Hflav}) cannot be further reduced as
for  Eq.(\ref{H12}), due to the non-trivial
time dependence of the flavor ladder operators.

In conclusion we find:
\bea\lab{Hineq} && \hspace {-.3cm} H_\sigma(t)\neq G_\theta^{-1}(t)  H_j
G_\theta(t)\; ,\;  \,(\sigma,j) = (e,1),(\mu,2).
\\ [3mm]\lab{Htot}
&&\hspace {-.3cm} \sum_\si  H_\si(t) =
\sum_j  H_j =  H\quad;\quad [H,G_\theta(t)]
\neq 0\,. \eea

The inequality sign in (\ref{Hineq}) can be  understood by noting
the appearance of the time derivative in the definition
(\ref{Hflav}) and the fact that the mixing generator is time
dependent. Consequently, the result (\ref{Htot}) is non-trivial and ensures
the fact that the expectation value of the total (flavor field)
energy on states in the flavor Hilbert space is time-independent.

We indeed have:
\bea\non {}_{e,\mu}\lan 0| H | 0 \ran_{e,\mu} & =& -\int
d^3{\bf k} \,(\om_{k,1} + \om_{k,2}) \times
\\ \lab{vevHf}&&\times(1\, - \, 2
\,\sin^2\te\,|V_k|^2 ) \eea
This has to be compared with the mass vacuum zero point
energy:
\bea\lab{vevHm} {}_{1,2}\lan 0| H | 0 \ran_{1,2} \, =\, -\int d^3{\bf k}
\,(\om_{k,1} + \om_{k,2}) \eea
The flavor vacuum zero-point energy has been studied in
Ref.\cite{cosmcost} in connection with
the cosmological constant.

Of course, both contributions Eqs.(\ref{vevHf}), (\ref{vevHm}) are
divergent and to properly define energy for flavor states, we need
to normal order the Hamiltonian with respect to the relevant
vacuum, namely the flavor vacuum:
\be \lab{nordf}
 \noH\, \equiv \,  H \, -\, {}_{e,\mu}\lan 0| H | 0 \ran_{e,\mu}
\ee
where the new symbol for the normal ordering was introduced to
remember that it refers to the flavor vacuum.

We finally obtain:
\bea \non
E_e(k) &\equiv & \lan \nu^{\bf k}_e| \noH
|\nu^{\bf k}_e\rangle \,
\\ [2mm]\label{dispe}  &=& \om_{k,1} \cos^2\te \,+ \,(1 -2
|V_{\bk}|^2)\,\om_{k,2} \sin^2 \te
\\ [3mm] \non
E_\mu(k) &\equiv& \lan \nu^{\bf k}_\mu| \noH |\nu^{\bf
k}_\mu\rangle \,
\\ [2mm] \label{dispmu} &=& \om_{k,2} \cos^2\te \,+ \,(1 -2
|V_{\bk}|^2) \,\om_{k,1} \sin^2 \te \eea
We propose to treat these as  modified dispersion relations and to
find the corresponding non-linear realization of the Lorentz
algebra as outlined in \cite{lorentz}. Obviously the energies in
Eqs.(\ref{dispe}),(\ref{dispmu}) are only  expectation values
subject to fluctuations but it is nevertheless sensible to
consider the modified Lorentz transformation for these dispersion
relations which form the classical limit of the theory.

Note the presence in the above dispersion relations, of the
Bogoliubov coefficient $|V_{\bk}|^2$: this term is due to the
flavor vacuum structure and is absent in the usual Pontecorvo
case. The maximum of the function $|V_{\bk}|^2$ occurs for
$k_{max}=\sqrt{m_1 m_2}$; we then have:
\bea |V_{k_{max}}|^2 = \frac{1}{2} -
\frac{1}{\sqrt{(1+\frac{m_1}{m_2})(1+\frac{m_2}{m_1})}} \eea
If we put $a=\frac{m_1}{m_2} < 1$, then the condition
$|V_{\bk}|^2 \ll \frac{1}{2}$ is realized for
\bea 1 > a > \frac{b^2 -1 +2 b (1 - \sqrt{2 b -1} )}{(b -1)^2}
\quad, \quad b\gg 1 \eea

For example, for $b=100$ we get $a<0.75$ which is compatible with
the current experimental bounds. This approximation was  used in
Ref.\cite{nulorentz} where for simplicity
we analyzed the  dispersion relations obtained from
the Pontecorvo states (\ref{Pont}). In the following we treat the full case
of Eqs.(\ref{dispe}),(\ref{dispmu}).

\section{Lorentz invariance for mixed neutrinos}

In this section, we study the dispersion relations (\ref{dispe}),
(\ref{dispmu}) and derive the corresponding non-linear realization
of the Lorentz algebra.

First, from Eqs.(\ref{dispe}), (\ref{dispmu}), let us define the
rest masses for the mixed neutrinos:
\bea\label{me} m_e &\equiv& E_e(k =0)\,=\, m_1 \cos^2\te + m_2
\sin^2 \te
\\ [2mm] \label{mmu}
m_\mu &\equiv& E_\mu(k =0)\,=\, m_2 \cos^2\te + m_1 \sin^2 \te
\eea

Then we investigate the high $k$ limit of
Eqs.(\ref{dispe}),(\ref{dispmu}). To first order in
$\frac{m_j^2}{2k}$, it is $\om_{k,j}\simeq k + \frac{m_j^2}{2k}$
and $|V_k|^2\simeq \frac{(m_2 -m_1)^2}{4 k^2}$ and we obtain:
\bea\non && E_e(k) \simeq k + \frac{{\wti
m}_e^2}{2k}\;; \quad {\wti m}_e^2 \equiv m_1^2 \cos(2\te) + m_1
m_2 \sin^2 \te
\\ [2mm]\non
&&E_\mu(k) \simeq k + \frac{{\wti m}_\mu^2}{2k}\;; \quad {\wti
m}_\mu^2 \equiv m_2^2 \cos(2\te) + m_1 m_2 \sin^2 \te \eea
where we introduced the effective masses ${\wti m}_e$ and ${\wti
m}_\mu$. We thus see that in
the high momentum (or equivalently high $E$ since it is a
monotonously growing function of $k$) limit, the dispersion
relations for the flavor neutrinos are indeed of the usual form, although
with a modified mass.

Noticing that $\om_{k,1}(1- 2 |V_\bk|^2) = \frac{k^2 + m_1
m_2}{\om_{k,2}}$ and $\om_{k,2}(1- 2 |V_\bk|^2) = \frac{k^2 + m_1
m_2}{\om_{k,1}}$, we rewrite Eqs.(\ref{dispe}), (\ref{dispmu}) as
\bea \non E_e(k) &=& \frac{2 k^2 + m_1 (m_2+m_1) - m_1
(m_2 -m_1) \cos(2\te)}{2 \sqrt{k^2 + m_1^2}}
 \\ \label{eqe1}
\\  [2mm]\non
E_\mu(k) &=& \frac{2 k^2 + m_2 (m_2+m_1) + m_2 (m_2 -m_1)
\cos(2\te)}{2 \sqrt{k^2 + m_2^2}}\\ \label{eqmu1}
 \eea
By introducing $a\equiv m_2/m_1 \ge 1$, we get
\bea \label{eqe2} E_e(k) &=& \frac{k^2 + m_1^2 -  (1 -a)
m_1^2\sin^2\te}{ \sqrt{k^2 + m_1^2}}
 \\ [2mm]\label{eqmu2}
E_\mu(k) &=& \frac{k^2 + a^2m_1^2 +
 a(1 -a) m_1^2\sin^2\te}{ \sqrt{k^2 + a^2 m_1^2}}
 \eea

It is easy to realize that, for $a>1$, the function $E_\mu(k)$ has
an absolute minimum at $k=0$ with the value $E_\mu(0)=m_\mu$.

The situation is different for $E_e(k)$: the minimum is now at
$k_{min} = \frac{1}{\sqrt{2}}\sqrt{a - 3 + (1-a)\cos(2\te)}$. This
is different from zero when $a$ is above the critical value
$a_c=\frac{\cos(2\te)-3}{\cos(2\te)-1}$. For $1<a<a_c$, the
function $E_e(k)$ has an absolute minimum at $k=0$ with the value
$E_e(0)=m_e$.

\vspace{.2cm}

This is represented in the two figures below for the case
$\te=\pi/6 \Lrar a_c =5$.

\begin{figure}[h]
\begin{center}
\includegraphics*[width=8.5cm]{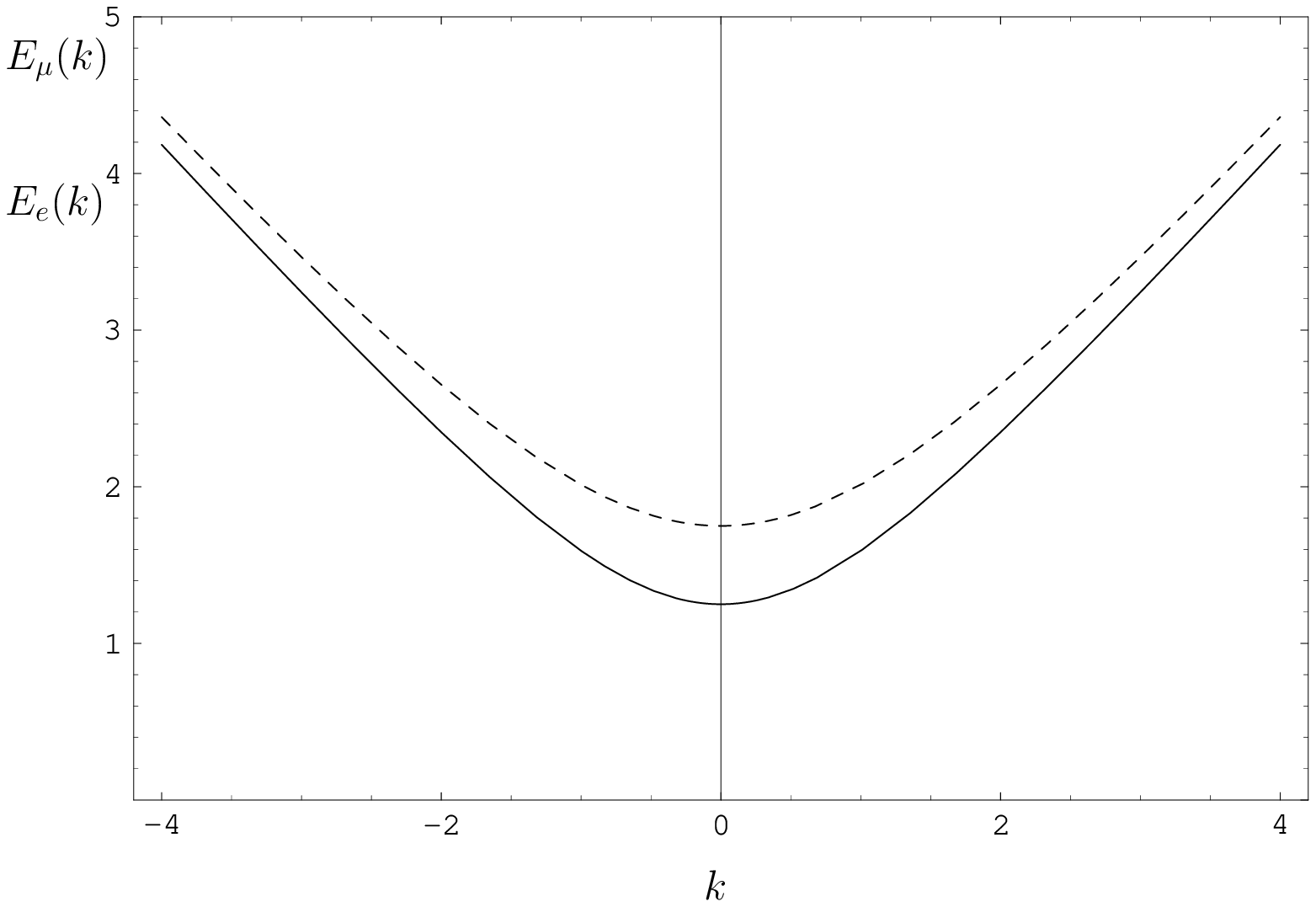}
\vspace{-0.5cm}
\end{center}
\caption{$E_e$ (solid line) and  $E_\mu$ (dashed line)
as functions of $k$   for $\te=\pi/6$, $m_\1=1$, $a=2$, $a_c=5$.} \label{fig1}
\end{figure}

\begin{figure}[h]
\begin{center}
\includegraphics*[width=8.5cm]{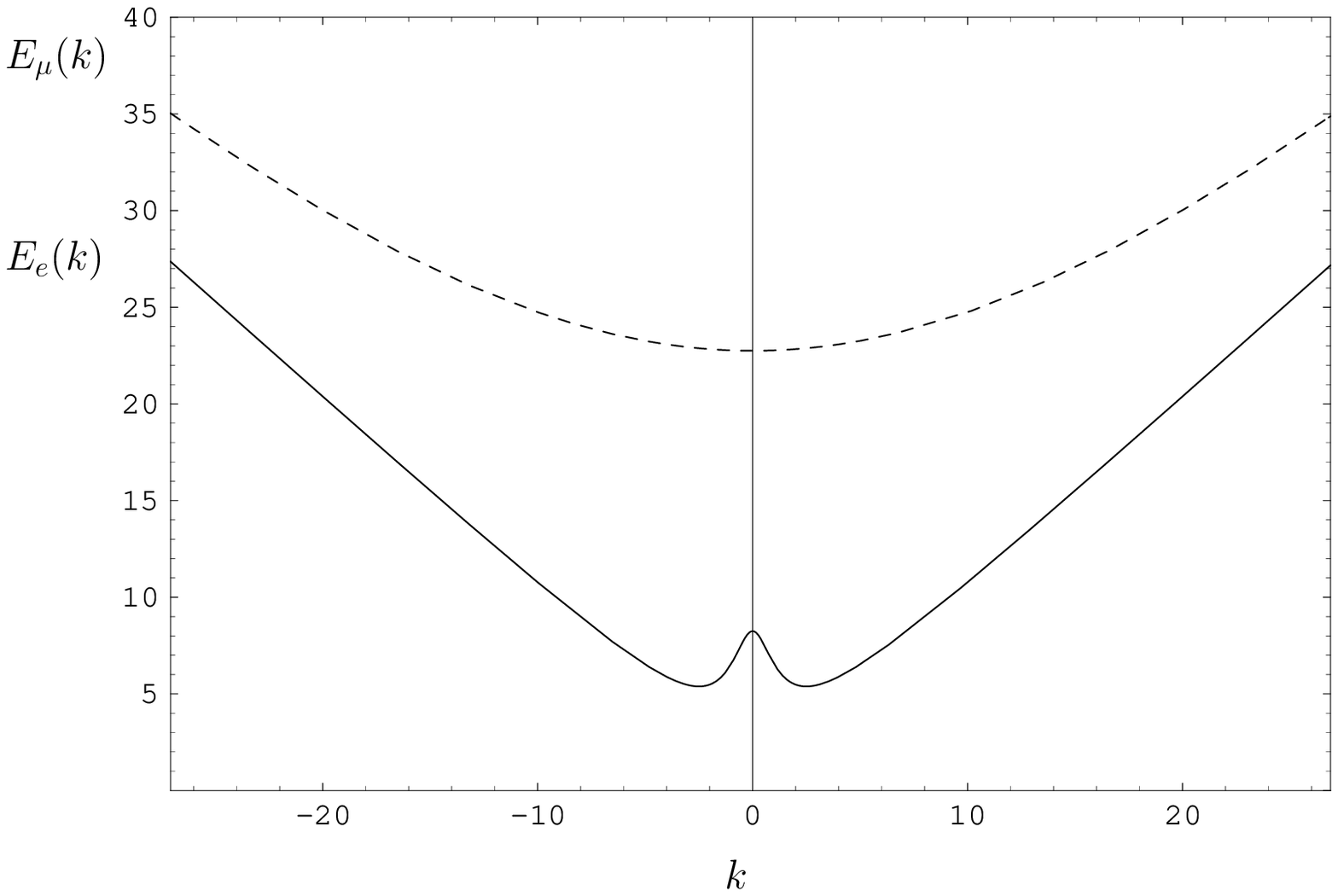}
\vspace{-0.5cm}
\end{center}
\caption{$E_e$ (solid line) and  $E_\mu$ (dashed line)
as functions of $k$   for $\te=\pi/6$, $m_\1=1$, $a=30$, $a_c=5$.} \label{fig2}
\end{figure}

In the following we consider only the subcritical case $a<a_c$.
The case $a>a_c$ will be treated elsewhere.

Following Ref.\cite{lorentz}, we now set the dispersion relations
in the following form:
\bea \label{lorentze} E_e^2 \, f_e^2(E_e)\, - \,k^2 \,g_e^2(E_e)\,
=\,  m_e^2
\\ [3mm] \label{lorentzmu}
E_\mu^2 \, f_\mu^2(E_\mu)\, -\, k^2\, g_\mu^2(E_\mu)\, =\,
m_\mu^2 \eea

It is now possible to identify the non-linear realization of the
Lorentz group which leaves these dispersion relations invariant.
They are generated by the transformation $ U \circ (E, {\bf
k})=(Ef,{\bf k}g) \label{udef} $ applied to the standard
Lorentz generators ($L_{ab} = p_a {\partial \over \partial p^b} -
 p_b {\partial \over \partial p^a}$):
\begin{equation}\label{U}
K^i = U^{-1} [p_0] L_0^{\ i} U [p_0 ]\, .
\end{equation}
This amounts to requiring linearity for the auxiliary variables ${\tilde
E}=Ef(E)$ and ${\tilde k}=kg(E)$. The resulting non-linear transformations
for $E$ and $k$ are a non-linear representation of the Lorentz group
ensuring that the deformed dispersion relations found for
flavor states are valid in all frames.

\vspace{.2cm}

We find:
\bea \non
f^2_e(E_e) &=&\frac{1} {2 E_e^2 \ } \Big[ 2
m_e(m_2-m_1) \sin^2\te
\\[2mm] \non
&&+E_e \left(E_e + \sqrt{E_e^2 - 4 m_1(m_2-m_1) \sin^2\te} \ri)\Big]
\\[2mm] \label{fge}
g_e(E_e) &=& 1
\eea
and \bea \non
f^2_\mu(E_\mu) &=&\frac{1} {2 E_\mu^2 \ } \Big[ -2
m_\mu(m_2-m_1) \sin^2\te
\\[2mm] \non
&&+E_\mu \left(E_\mu + \sqrt{E_\mu^2 + 4
m_2(m_2-m_1) \sin^2\te} \ri)\Big]
\\[2mm] \label{fgmu}
g_\mu(E_\mu) &=& 1
\eea
It is easy to check that, for $m_1=m_2$ and/or $\te=0$, we have
$f^2_e(E_e)=f^2_\mu(E_\mu)=1$. Also $f^2_\mu(m_\mu)=1$ (for any
$a\ge 1$) and $f^2_e(m_e)=1$ (only for $a_c\ge a\ge 1$)

\vspace{.2cm}

A plot of these two functions is given below:
\begin{figure}[h]
\begin{center}
\includegraphics*[width=8.5cm]{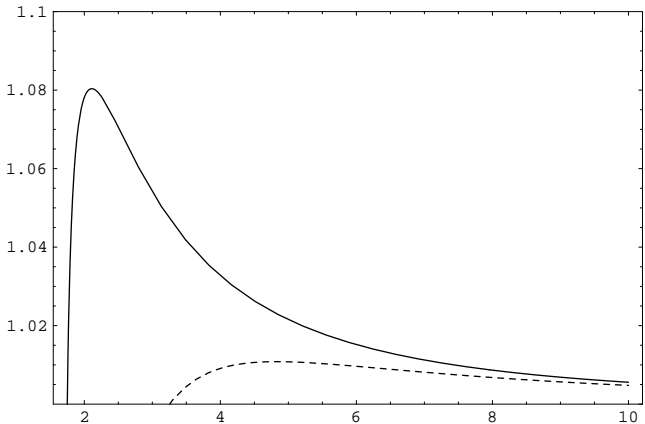}
\vspace{-0.5cm}
\end{center}
\caption{$f_e$ as a function of $E_e$ (solid line) and
$f_\mu$ as a function of $E_\mu$ (dashed line) for $\te=\pi/6$,
$m_1=1$, $a=4$, $a_c=5$.} \label{fig3}
\end{figure}

For large values of $E_e$ and $E_\mu$, $f^2_e(E_e)$ and
$f^2_\mu(E_\mu)$ can be approximated as: \bea
f^2_e(E_e)\,\simeq\,1 \, -\, \frac{1}{ E_e^2} (m_2 -m_1)^2
\sin^4\te
\\
f^2_\mu(E_\mu)\,\simeq\,1 \, -\, \frac{1}{ E_\mu^2} (m_2 -m_1)^2
\sin^4\te
\eea
showing that the Lorentzian regime is
approached quadratically as the energy (momentum) grows.

\section{Phenomenological consequences}

We now focus on some phenomenological
consequences which arise from considering the flavor states as
fundamental and consequently the non-standard dispersion relations (\ref{dispe}),
(\ref{dispmu}) as characterizing mixed neutrinos.

We consider the case of a beta decay process like tritium decay, which allows for
a direct investigation of neutrino mass. In the following we take into account
the various possible outcomes of this experiment in correspondence of the
different theoretical possibilities for the nature of mixed neutrinos.
We show that significative differences arise at phenomenological level
between the standard theory and the scenario above described.

Let us then consider the decay:
\bea\non A\rightarrow B + e^- +{\bar \nu}_e \eea
where $A$ and $B$ are two nuclei (e.g. $^3$H and $^3$He).

The electron  spectrum is proportional to phase volume factor $E p
E_{e} p_{e}$:
\bea \label{specpref} \frac {dN}{dK}= C E p\,
(Q-K)\sqrt {(Q-K)^2\,-\,m_{e}^2} \eea
where $E=m+K$  and $p=\sqrt{E^2-m^2}$  are electron's
energy and momentum.
We denote by $m_e$ the electron (anti-)neutrino mass.

The endpoint of $\beta$ decay is the  maximal kinetic energy $K_{max}$
the electron can take (constrained by the available energy
$Q=E_A-E_B-m\approx m_A-m_B-m$). In the case of  tritium decay, $Q=18.6$ KeV.
$Q$ is shared between the
(unmeasured) neutrino energy and the (measured) electron kinetic
energy $K$.

It is clear that if  the neutrino were massless, then $m_e=0$ and $K_{max}=Q$.

On the other hand, if the neutrino were a mass eigenstate (say with $m_e=m_1$),
then $K_{max}=Q-m_1$.

\vspace{.5cm}

 We now consider the various possibilities which can arise
in the presence of  mixing:

\vspace{.2cm}

$\bullet\;${If, following the common wisdom, mass eigenstates are considered
fundamental, the} $\beta$ { spectrum is}
\bea \label{specmass}
\frac {dN}{dK}= C E p \,E_e \sum_j |U_{ej}|^2 \sqrt{E_e^2-m_{j}^2}\;
\Theta (E_e-m_j)
\eea
where $E_e=Q-K$ and
$U_{ej}=(\cos\theta,\sin\theta)$ and $\Theta (E_e-m_j)$ is the Heaviside step function.

The end point is at $K=Q-m_1$  and the spectrum has an
inflexion at $K\simeq Q-m_2$.

\vspace{.4cm}

If flavor neutrinos are to be taken as fundamental, we have the following two
options:

\vspace{.2cm}

$\bullet\;$  Assuming that nuclei and the electron satisfy linear
Lorentz transformations, and that $E_{e}f_e(E_{e})$  transforms
linearly, the only covariant law of energy conservation is
\bea \non E_A = E_B + E + E_{e} f_e(E_{e})\, . \eea

The endpoint of  $\beta$ decay is now $ K_{max}=Q-m_e$  and the
$\beta$ spectrum is proportional to the phase volume factor
$E p E_{e}f_e(E_{e}) p_{e}$:
\bea
\frac {dN}{dK}=C E p\, (Q-K)\sqrt {(Q-K)^2-m_{e}^2}\;\Theta
(E_e-m_e)\ \
 \eea

\vspace{.2cm}

$\bullet\;$ If, on the contrary, we insist upon the standard law
\bea E_A = E_B + E + E_{e} \non \eea
we have introduced a preferred
frame, and are in conflict with the principle of relativity.

Then $K_{max}=Q- m_{e}$   and the spectrum is proportional to
the phase volume factor $E p E_{e}p_{e}$:
\bea \non
\frac {dN}{dK}= C E p\, (Q-K)\sqrt {(Q-K)^2f_e^2\,-\,m_{e}^2}\; \Theta
(E_e-m_e)\\ \label{specpref}
\eea
%

\begin{widetext}

\begin{figure}
\begin{center}
\includegraphics*[width=14.5cm]{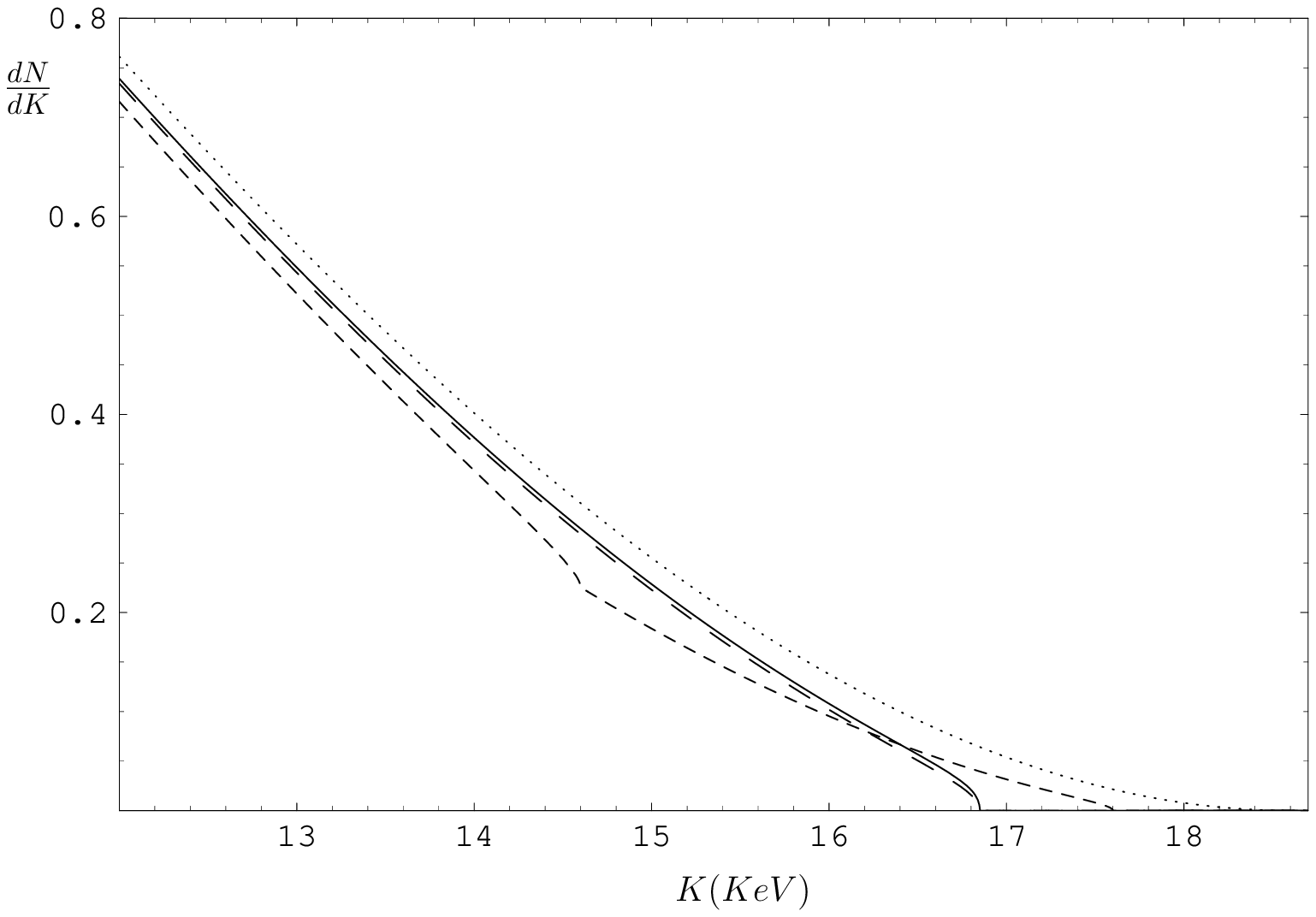}
\vspace{-0.5cm}
\end{center}
\caption{The tail of the tritium $\beta$ spectrum for:
- a massless neutrino (dotted line);
- a Lorentz invariant flavor state  (solid line);
- preferred frame (long-dashed line);
-  superposed prediction for 2 mass states (short-dashed line):
notice the inflexion in the spectrum where the most massive state
switches off.
We used $m_e=1.75$ KeV, $m_1=1$ KeV, $m_2=4$
KeV, $\theta=\pi/6$.} \label{fig4}
\end{figure}

\end{widetext}

The above possibilities are plotted in Fig.(\ref{fig4}), together
with the spectrum for a massless neutrino, for comparison.

We note that the next generation tritium beta decay experiments will allow a sub-eV
sensitivity for the electron neutrino mass \cite{katrin}, thus hopefully allowing to
unveil the true nature of mixed neutrinos.

\section{Conclusions}

In this paper, we have investigated some aspects of neutrino mixing in Quantum
Field Theory. From a careful analysis of the Hilbert space structure
for flavor (mixed) fields
it has emerged that the flavor states, defined as
eigenstates of the flavor charge, are at odds
with Lorentz invariance. Indeed they exhibit
non-standard dispersion relations, which however
reduce to the usual (Lorentzian) ones in the relativistic limit.

\vspace{.2cm}

We have then shown that it is possible to account for
such a modified dispersion relations, by
resorting to a recent proposal \cite{lorentz}:
According to this, we could identify a non-linear
representation of the Lorentz group allowing
for these dispersion relations and ensuring at the
same time the equivalence of inertial observers.

\vspace{.2cm}

Finally, we have considered possible phenomenological
consequences which can arise from our analysis,
by looking at the beta decay. We have considered
various possibilities, including that of introducing a
preferred frame, and shown that observable
differences arise in correspondence of  the various cases.

\acknowledgments

M.~B. acknowledges partial support from  MURST, INFN, INFM and ESF Program COSLAB.
P.~P.~P. thanks the FCT (part of the Portuguese Ministry of Education)
for financial support under scholarship SFRH/BD/10889/2002.

\appendix

\section{Flavor Hilbert space}

The free fields $\nu_{1}(x)$ and $\nu_{2}(x)$ are written as
\bea \non \nu_j(x)& = &\sum_{r=1,2} \int
\frac{d^3\bk}{(2\pi)^\frac{3}{2}}  e^{i {\bf k}\cdot {\bf x}}
\lf[u^{r}_{{\bf k},j}(t) \al^{r}_{{\bf k},j}\:
\ri.
\\ \lab{nui} &&\lf.
+ v^{r}_{-{\bf
k},j}(t) \bt^{r\dag }_{-{\bf k},j} \ri], \qquad  \;j=1,2.\eea
Here $u^{r}_{{\bf k},j}(t)=e^{-i\om_{k,j} t}u^{r}_{{\bf k},j}$
and $v^{r}_{{\bf k},j}(t)=e^{i\om_{k,j} t}v^{r}_{{\bf k},j}$, with
$\om_{k,j}=\sqrt{|\bk|^2+m_j^2}$.

The orthonormality and
completeness relations are:
\bea \non && u^{r\dag}_{{\bf k},j} u^{s}_{{\bf k},j} = v^{r\dag}_{{\bf
k},j} v^{s}_{{\bf k},j} = \de_{rs} \;,
\\ [2mm] \non
&&u^{r\dag}_{{\bf k},j} v^{s}_{-{\bf k},j} = v^{r\dag}_{-{\bf k},j} u^{s}_{{\bf
k},j} = 0\;, \\[2mm]
&&\sum_{r}(u^{r}_{{\bf k},j}u^{r\dag}_{{\bf k},j}  + v^{r}_{-{\bf k},j}v^{r\dag}_{-{\bf
k},j} ) = \ide_2 \;. \eea where $\ide_n$ is the $n \times n$ unit
matrix.

The $\al^{r}_{{\bf k},j}$ and
the $\bt^{r }_{{\bf k},j}$, $j,r=1,2$ are the annihilation
operators for the vacuum state
$|0\ran_{1,2}\equiv|0\ran_{1}\otimes \,|0\ran_{2}$: $\al^{r}_{{\bf
k},j}|0\ran_{12}= \bt^{r }_{{\bf k},j}|0\ran_{12}=0$.

 The
anticommutation relations are:
\bea\non &&{}\hspace{-.5cm}
\{\nu^{\al}_{i}({\bf x}), \nu^{\bt\dag }_{j}({\bf
y})\}_{t=t'} =  \, \de^{3}({\bf{x}}-{\bf{y}}) \de_{\al\bt}
\de_{ij} \;, \;\; \al,\bt=1,..,4 ,
\\ \non
&&{}\hspace{-.5cm}
\{\al^{r}_{{\bf k},i}, \al^{s\dag }_{{\bf q},j}\} = \de^3 ({\bk
- \bp})\de _{rs}\de_{ij} \, ,
 \\ &&{}\hspace{-.5cm}\{\bt^{r}_{{\bf k},i},
\bt^{s\dag}_{{\bf q},j}\} = \de^3({\bk - \bq})
\de_{rs}\de_{ij},\;\;\;\; i,j=1,2\;. \eea
All other anticommutators are zero.

In the reference frame where ${\bf{k}}$ is collinear with ${\hat
{\bf k}} \equiv (0,0,1)$, the flavor annihilation operators have the
simple form:
\bea \non\al^{r}_{{\bf k},e}(t)&=&\cos\te\;\al^{r}_{{\bf
k},1}\;+\;\sin\te\;\lf( U_{{\bf k}}^{*}(t)\; \al^{r}_{{\bf
k},2}\;\ri.
\\
&&\lf.+\;\ep^{r}_{\bf k}\;
V_{{\bf k}}(t)\; \bt^{r\dag}_{-{\bf k},2}\ri) \\
\non
\al^{r}_{{\bf k},\mu}(t)&=&\cos\te\;\al^{r}_{{\bf
k},2}\;-\;\sin\te\;\lf( U_{{\bf k}}(t)\; \al^{r}_{{\bf
k},1}\;\ri.
\\
&&\lf.-\;\ep^{r}_{\bf k}\;
V_{{\bf k}}(t)\; \bt^{r\dag}_{-{\bf k},1}\ri)
\\
\non
\bt^{r}_{-{\bf k},e}(t)&=&\cos\te\;\bt^{r}_{-{\bf
k},1}\;+\;\sin\te\;\lf( U_{{\bf k}}^{*}(t)\; \bt^{r}_{-{\bf
k},2}\;\ri.
\\
&&\lf.-\;\ep^{r}_{\bf k}\; V_{{\bf k}}(t)\; \al^{r\dag}_{{\bf
k},2}\ri)
\\
\non \bt^{r}_{-{\bf k},\mu}(t)&=&\cos\te\;\bt^{r}_{-{\bf
k},2}\;-\;\sin\te\;\lf( U_{{\bf k}}(t)\; \bt^{r}_{-{\bf
k},1}\;\ri.
\\
&&\lf.+\;\ep^{r}_{\bf k}\; V_{{\bf k}}(t)\; \al^{r\dag}_{{\bf
k},1}\ri) \eea
where $\ep^{r}_{\bf k}\equiv (-1)^{r + {\bf k}\cdot {\hat {\bf k}}
+ 1}$  and $U_{{\bf k}}(t)$, $V_{{\bf k}}(t)$ are Bogoliubov
coefficients given by:
\bea \non U_{{\bf k}}(t)&\equiv& u^{r\dag}_{{\bf k},2}(t)u^{r}_{{\bf
k},1}(t)= v^{r\dag}_{-{\bf k},1}(t)v^{r}_{-{\bf k},2}(t)
\\ \non
&=& |U_{{\bf k}}|\;e^{i(\om_{k,2}-\om_{k,1})t} \,,
\\ [2mm]\non
V_{{\bf k}}(t)&\equiv& \ep^{r}_{\bf k}\;u^{r\dag}_{{\bf
k},1}(t)v^{r}_{-{\bf k},2}(t)= -\ep^{r}_{\bf k}\;u^{r\dag}_{{\bf
k},2}(t)v^{r}_{-{\bf k},1}(t)
\\
&=&|V_{{\bf k}}| \;e^{i(\om_{k,2}+\om_{k,1})t} \,,
\eea
with
\bea \non
&& |U_{{\bf k}}|=\frac{|{\bf k}|^{2} +(\om_{k,1}+m_{1})(\om_{k,2}+m_{2})}{2
\sqrt{\om_{k,1}\om_{k,2}(\om_{k,1}+m_{1})(\om_{k,2}+m_{2})}} \,,
\\ \non &&
|V_{{\bf k}}|=\frac{ (\om_{k,1}+m_{1}) - (\om_{k,2}+m_{2})}{2
\sqrt{\om_{k,1}\om_{k,2}(\om_{k,1}+m_{1})(\om_{k,2}+m_{2})}}\, |{\bf k}| \, ,
\\ \mlab{2.40}
&&|U_{{\bf k}}|^{2}+|V_{{\bf k}}|^{2}=1.
\eea
%

\bibliography{apssamp}

\end{document}